\documentclass{article}
\usepackage{spconf,amsmath,graphicx}
\usepackage{multirow}
\newcommand{\tildeapprox}{\raisebox{0.5ex}{\texttildelow}}
\usepackage{threeparttable}

\title{Deep Transformer based Data Augmentation with Subword Units for Morphologically Rich Online ASR}
\name{Balázs Tarján$^{1,2}$, György Szaszák$^1$, Tibor Fegyó$^{1,2}$, Péter Mihajlik$^{1,3}$}
\address{
  $^1$Budapest University of Technology and Economics, Budapest, Hungary\\
  $^2$SpeechTex Ltd., Budapest, Hungary\\
  $^3$THINKTech Research Center, Vác, Hungary}  

\begin{document}
%
\maketitle
\begin{abstract}
Recently Deep Transformer models have proven to be particularly powerful in language modeling tasks for ASR.
Their high complexity, however, makes them very difficult to apply in the first (single) pass of an online system.
Recent studies showed that a considerable part of the knowledge of neural network Language Models (LM) can be transferred to traditional n-grams by using neural text generation based data augmentation.
In our paper, we pre-train a GPT-2 Transformer LM on a general text corpus and fine-tune it on our Hungarian conversational call center ASR task.
We show that although data augmentation with Transformer-generated text works well for isolating languages, it causes a vocabulary explosion in a morphologically rich language.
Therefore, we propose a new method called subword-based neural text augmentation, where we retokenize the generated text into statistically derived subwords.
We compare Morfessor and BPE statistical subword tokenizers and show that both methods can significantly improve the WER while greatly reducing vocabulary size and memory requirements.
Finally, we also demonstrate that subword-based neural text augmentation outperforms the word-based approach not only in terms of overall WER but also in recognition of OOV words.

\end{abstract}
\section{Introduction and problem statement}
\label{sec:intro}

Today's state-of-the-art in language modeling for ASR relies on neural Language Models (LM)~\cite{Irie2019,mikolov2010recurrent,sundermeyer2012lstm},
capable of handling continuous space and thereby outperforming traditional Back-off N-gram LMs (BNLM).
BNLMs cannot exploit long context based syntactic dependencies and are also less flexible in terms of generalization for unseen cases,
as semantic knowledge (such as embeddings reflecting similarity) is not captured while training them. 

Neural LMs however have an undesired property, they are computationally very heavy in decoding,
so neural LMs cannot be effectively used in a single decoding pass,
they are rather exploited by rescoring lattices obtained from a first decoding pass with a BNLM.
It is obvious, but can also be shown, that information is lost during the first decoding pass,
as the pruning of the recognition network is based only on short context syntax,
discarding both longer context syntactic and quasi all semantic knowledge.
Another problem arising is the increased latency of the system through the two decoding passes,
which hampers exploitation in strict online requirements.

To reduce these limitations in exploiting neural LMs for ASR,
several solutions have been proposed~\cite{Singh2019,Deoras2011,Arisoy2014,Adel2014}.
In~\cite{Adel2014} it was shown that using the neural LM to generate an augmented training corpus to train an improved BNLM is the best performing strategy.
Such a BNLM trained on augmented corpus can be used in a single pass or in the first pass of decoding.
Sometimes these are called approximative models as they try to capture
the knowledge of the neural model through their augmented training corpus.
Suzuki et al.~\cite{Suzuki2019} uses a domain balanced mixture of the training corpora to train a shallow RNNLM for text generation,
and improve speech recognition results for Japanese, Korean and English.
Wang et al.~\cite{Wang2019a} report using general domain pre-trained Transformer~\cite{Vaswani2017} to augment text corpora used to train LMs.
They demonstrate that the pre-trained and fine-tuned Transformer performs significantly better in data augmentation than LSTMs or simple in-domain Transformer models.

Another burden of language modeling for morphologically rich languages are the different syntactic properties of the language compared to English.
Heavy agglutination results in much larger vocabularies, which is a problem in itself, but causes other problems too:
individual word forms occur less often and hence,
the size of the training corpus should accordingly be augmented to maintain the predictive power of the dataset.
Moreover, as suffixes express grammatical relations usually provided by word order in English,
morphologically rich languages tend to be more permissive in choosing word order,
leading to higher variation.
This impairs BNLM estimation badly, but may also cause that word embeddings become less powerful in terms of syntactic and semantic consistency~\cite{Dobrossy2019},
even despite using long context windows.

To alleviate these problems linked to the different organization of morphologically rich languages,
subword unit modeling is an often used alternative.
Subword unit based ASR has been demonstrated to improve WER for several morphologically rich languages~\cite{Creutz2002,Kurimo2006487}.
Suzuki et al.~\cite{Suzuki2019} use subword approach for data augmentation
to enrich text corpora to train BNLM,
but compose these subwords back into words to prepare the final LM, unlike our approach that retokenizes words into subword units in the final LM. 
 
In this paper we aim to improve LM for an online call center ASR system in the morphologically rich Hungarian.
We use parliamentary text to pre-train a GPT-2 structure Transformer LM~\cite{Radford2019},
and fine-tune it on the target domain.
With this model we generate training text for a BNLM.
We demonstrate that such Transformer based data augmentation is efficient in morphologically rich Hungarian,
if vocabulary is large enough and a large BNLM is used.
Retokenizing the augmented training corpus to subword units,
and training a subword-based BNLM on it, we demonstrate that
(i) the ASR accuracy further improves compared to the word based baseline augmented BNLM,
and (ii) the footprint and complexity of the resulting subword unit augmented BNLM significantly decrease.
As subword unit LMs are known to perform better on a wide range of morphologically rich languages,
we hypothesize that our approach is transferable to other such languages.
We consider as novelties of our paper the following:
(i) we propose the retokenization of the Transformer augmented LM training corpus; 
(ii) we are the first to use the GPT-2 Transformer structure to augment LM training corpora;
(iii) we are the first to apply a Transformer based LM for a Hungarian ASR task;
(iv) we demonstrate that the subword-based neural text augmentation can be exceptionally efficient in modeling OOV words.

\section{Datasets and models} 

\subsection{Datasets} \label{database}


In-domain training data is extracted from the Hungarian Call Center Speech Database (HCCSD) consisting of anonymised telephone customer service calls and the corresponding manual transcripts.
We selected 290 hours of recordings for training the acoustic model of our ASR system (see Table~\ref{data_stat}).
The in-domain LMs are built on the transcripts of the training set containing 3.4M word tokens and 100k unique word forms.
As the available in-domain training text data is very limited, we also utilize a general text corpus for pre-training the Transformer LM,
which was collected from the website of the Hungarian National Assembly\footnote{www.parlament.hu} and contains official transcripts of parliamentary speeches.

\renewcommand{\arraystretch}{1.1}
\begin{table}[tbp]
\centering
\caption{Train and test dataset statistics}
\begin{tabular}{rccc}
\hline
 \textbf{In-domain}                  & Train & Valid. & Eval. \\ \hline
Audio {[}h:m{]} & 290:07 & 7:31                & 12:12               \\
\# of word tokens       & 3,401,775 & 45,773               & 66,312               \\
word OOV rate {[}\%{]}  & -- & 2.7                 & 2.5                 \\ \hline
\textbf{General text}  & & & \\ 
\# of word tokens       & 57,601,277 & --               & --              \\ \hline
\end{tabular}
\label{data_stat}
\end{table}


For testing purposes another 20 hours of transcribed conversations are selected from the HCCSD and is split into two disjoint sets (see Table~\ref{data_stat}).
The validation set (\tildeapprox 7.5 hours) is used for hyperparameter optimization for the tokenizers and the language models.
The evaluation set (\tildeapprox 12 hours) is reserved to compare the performance of the different modeling approaches addressed in this paper (see Sec.~\ref{results}).


\subsection{Back-off n-gram language models}

Count-based, back-off language models (BNLMs) have low computational cost and fit well into Weighted Finite-State Transducer (WFST) framework, hence are still widely used in online, single-pass ASR systems.
We carry out training and interpolation of BNLMs with the SRI language modeling toolkit~\cite{Stolcke2002} applying Chen and Goodman's modified Kneser-Ney discounting~\cite{Chen1999}.
After experimenting with different n-gram orders on the development set, we found 4-gram models the optimal choice both for word and subword BNLMs.

\subsection{Transformer language model} \label{transformer}


Recently Transformer architectures have proven to be particularly successful in generating well-structured, high-quality texts thanks to the self attention mechanism and the depth of the model~\cite{Radford2019,Yang2019}.
In order to generate augmentation text to our ASR task, we applied one of the most promising Transformer architectures called OpenAI GPT-2~\cite{Radford2019} implemented in HuggingFace's Transformers library~\cite{Wolf2019HuggingFacesTS}.
The GPT-2 architecture has four variants with different sizes from which we opted for the \textit{medium} using a total of 345M parameters.

\begin{figure}[tbp]
\centering
\includegraphics[width=0.64\columnwidth]{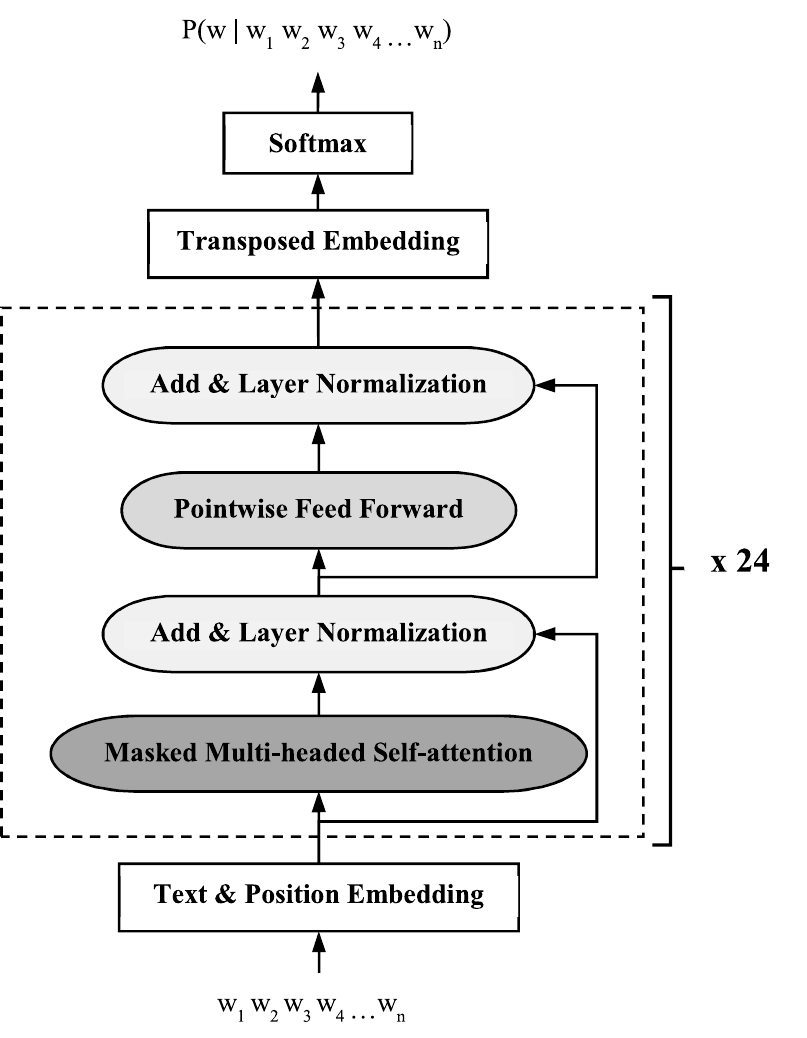}
\caption{Structure of the applied GPT-2 \textit{medium} architecture}
\label{fig:gtp2_structure}
\end{figure}

GPT-2 \textit{medium} consists of 24 decoder-only Transformer blocks each having 16 attention heads and 1024 dimensional embedding and bottleneck layers (see Fig.~\ref{fig:gtp2_structure}).
Due to GPT-2 conventions the first feed forward layer in each block is four times larger than the bottleneck layers (4096).
For regularization purposes it applies embedding, attention and residual dropouts with a rate of 0.1.
We apply the Adam optimization scheme~\cite{Kingma2015} with initial learning rate of 1e-4 and a linear decay schedule.
We pre-train the model on the general training corpus for 15 epochs using minibatches of 16 sequences consisting of 512 tokens each.
Fine-tuning of the pre-trained model took 4 epochs on the in-domain training set with the same hyperparameters as in pre-training.
Tokenization was performed with a byte-level Byte Pair Encoding (BPE)~\cite{Sennrich2015} model with 30k vocabulary items (256 bytes + 29744 merges) trained on the in-domain training set.

\section{Data augmentation} \label{data_augmentation}

\subsection{Neural text generation}

Generation of a text sequence is initialized with a prefix prompt, which we sample from the in-domain training set.
The length of the sampled prefix varies randomly between $1..7$ words to balance the trade-off between free and constrained text generation.
For the same reasons, the temperature is also randomly changed from 1.0 to 1.5.
We generate two large corpora (1 billion words each) for data augmentation purposes.
One with the pre-trained and then fine-tuned Transformer LM (TR) described in Sec.~\ref{transformer} and one with a Transformer trained directly on the in-domain corpus without pre-training (TR-noPT).

\subsection{Word-based data augmentation}

The fact that text corpora generated by RNNLMs can improve the accuracy of n-gram language models has been shown by several studies before~\cite{Deoras2011,Adel2014,Suzuki2019,Tarjan2019a}.
However, Wang and her colleagues~\cite{Wang2019a} were the first who applied general domain pre-training and in-domain fine-tuning of a Transformer LM to improve the effectiveness of the data augmentation process.
For that reason we summarize their original, word-based data augmentation process in this section.
In the next section, we are going to propose an extended version of the augmentation process that fits better to morphologically rich scenarios. 

\begin{figure}[tbp]
\centering
\includegraphics[width=1.0\columnwidth]{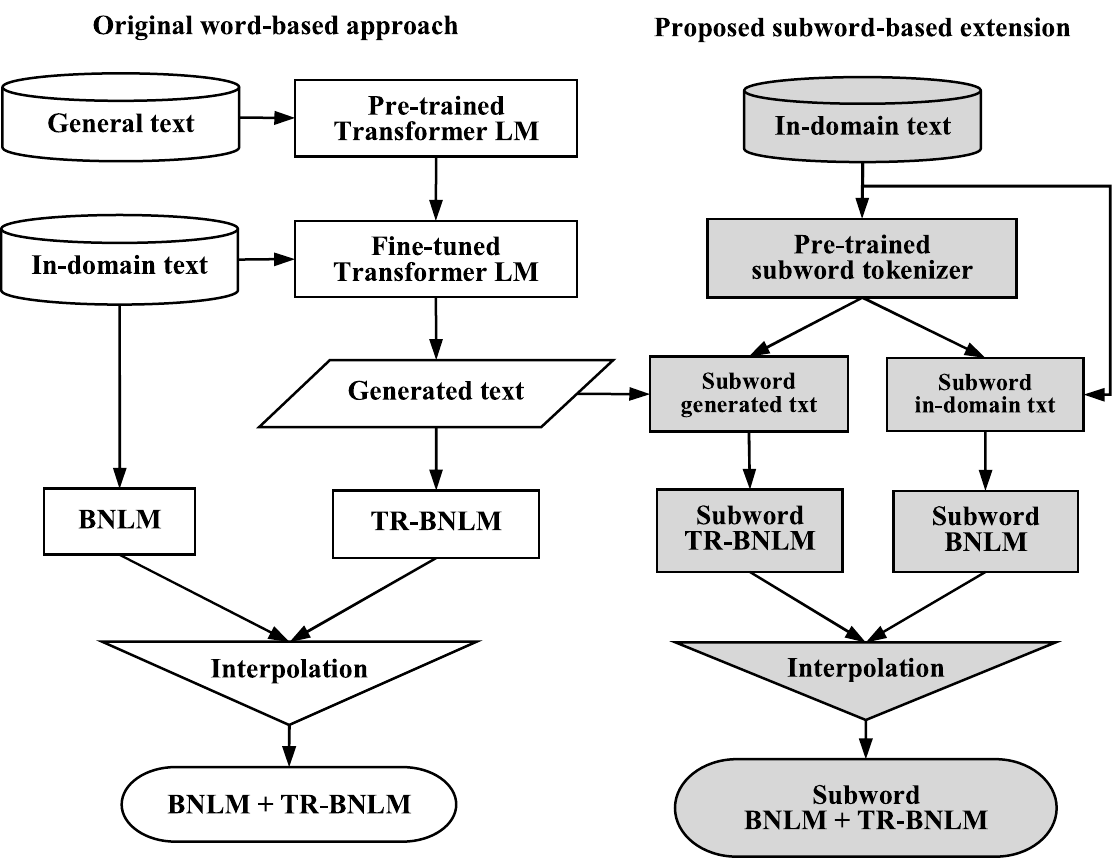}
\caption{Neural text generation based data augmentation of language models with the proposed modification (gray boxes)}
\label{fig:augmentation}
\end{figure}

The original, word-based version of neural text based data augmentation process is shown on the left side of Fig.~\ref{fig:augmentation} (white boxes).
First a large corpus is generated by the Transformer LM (pre-trained on a general text corpus and fine-tuned on the in-domain text).
Based on this generated text a BNLM (TR-BNLM) is trained, which approximates the short-term dependencies learned by the Transformer.
To further improve the model, the TR-BNLM can be interpolated with a BNLM trained the on the in-domain text (BNLM + TR-BNLM).


\subsection{The proposed subword-based data augmentation} \label{sec:subword_aug}

Morphologically rich languages have significantly larger vocabulary, as case endings usually reflect grammatical roles.
Large vocabulary size can be a problem in itself, however it also increases data sparseness in the training data.
A common remedy is to segment words into smaller units and train language models on these subword sequences~\cite{Kurimo2006487,Mihajlik2010}.
In this paper we compare two popular, data-driven subword tokenizers Morfessor~\cite{Creutz2002} and BPE~\cite{Sennrich2015} algorithm.
Morfessor is inspired by the Minimum Description Length (MDL) principle and specifically designed for processing morphologically rich languages. 
We apply its Python implementation called Morfessor 2.0~\cite{Virpioja2013a}.
BPE is one of the most widely used subword tokenizers especially for processing byte-level token sequences~\cite{Radford2019}.
Using byte-level tokens, however, could cause ambiguity in the phonetic transcription of multi-byte characters of Hungarian, hence both tokenizers are applied on the level of characters.





Our proposed extension to the word-based data augmentation process called \textit{subword-based neural text augmentation} is depicted in Fig.~\ref{fig:augmentation} (grey boxes).
The revised data augmentation process starts with training the subword tokenizer (Morfessor or BPE) on the in-domain dataset.
The word-based generated text corpus and the in-domain training text are then segmented into subword sequences using the pre-trained tokenizer.
In order to preserve word boundary information during the ASR decoding process, non-initial subwords were tagged with the `+' sign.
For instance, subword segmentation transforms the Hungarian sentence `\textit{megbeszélem a nejemmel}' (meaning `I will discuss it with my wife') as follows: `\textit{meg +beszél +em a nejem +mel}'.
Based on the segmented text, we train BNLM models (Subword BNLM and Subword TR-BNLM in Fig.~\ref{fig:augmentation}), which can be interpolated for the best performance again (Subword BNLM + TR-BNLM).


\section{Results and Discussion} \label{results}


\subsection{ASR setup}

High resolution MFCC vectors were used as input features for an LF-MMI trained Factored Time Delay Neural Network (TDNN-F) acoustic model~\cite{Povey2018,Povey:192584}.
The matrix size (hidden-layer dimension) was 768 and the linear bottleneck dimension was 80 resulting in a total number of 6M parameters in the 12 hidden layers.
Phoneme-based acoustic and language model resources were compiled into WFSTs and decoded with the VoXerver~\cite{5999466} ASR decoder.
The typical latency of the online decoding setup was measured to be around 250 ms.



\begin{table}[tbp]
\centering
\caption{WER of the baseline and neural augmented language models using word-based modeling and 100k vocabulary}
\begin{tabular}{lcc}
\hline
\textbf{Model}            & \textbf{\begin{tabular}[c]{@{}c@{}}WER\\ {[}\%{]}\end{tabular}} & \textbf{\begin{tabular}[c]{@{}c@{}}WERR\\ {[}\%{]}\end{tabular}} \\ \hline
BNLM                      & 21.9                                                            & -                                                                \\
RNN-BNLM                  & 22.5                                                            & -2.6*                                                            \\
TR-noPT-BNLM              & 23.1                                                            & -5.3*                                                            \\
TR-BNLM                   & 21.5                                                            & 1.7*                                                             \\ \hline
BNLM + RNN-BNLM           & 21.3                                                            & 2.7*                                                             \\
BNLM + TR-noPT-BNLM       & 21.1                                                            & 3.7*                                                             \\
BNLM + TR-BNLM            & 20.6                                                            & 5.9*                                                             \\ \hline
BNLM + TR-BNLM + RNN-BNLM & 20.4                                                            & 6.8*                                                             \\ \hline
\end{tabular}
\begin{tablenotes}
\item * sign indicates significant difference compared to BNLM and was tested with Wilcoxon signed-rank test (p \textless~0.05).
\end{tablenotes}
\label{table:wer_word}
\end{table}

\subsection{Results with word-based augmentation}

\subsubsection{Comparing language modeling approaches} \label{label:results_word_100k}

In our first experiment, we use the augmentation text in its original form without subword segmentation.
Our goal is to compare the modeling capabilities of language models on in-vocabulary words, hence we limited the vocabulary of all models to the 100k words occurring in the in-domain training text.
In our previous work~\cite{Tarjan2019a}, we augmented the same dataset with a corpus generated by a 2-layer LSTM RNNLM.
The results from this former study are placed in this paper to serve as an advanced baseline (RNN-BNLM).
All models were pruned to 1~GB runtime memory usage.
The results are summarized in Table~\ref{table:wer_word}.

Without LM interpolation, neither the RNN-BNLM nor the TR-noPT-BNLM (Transformer without pre-training) models can outperform the baseline BNLM.
Only the pre-trained TR-BNLM can reduce the Word Error Rate (WER) by around 2\% relative.
In contrast, with LM interpolation, all augmentation methods reduce significantly the WER of the baseline model.
Using a recurrent model (BNLM + RNN-BNLM) or the non-pretrained Transformer (BNLM~+ TR-noPT-BNLM) for data augmentation result in similar Word Error Rate Reduction (WERR), with the Transformer model being slightly better (2.7\% vs. 3.7\% WERR).
The pre-trained Transformer (BNLM + TR-BNLM), however, stands out among all other approaches, since it reduces the error rate by relative 6\%.
We also tested whether augmentation models can support each other and found that by applying RNN-BNLM and TR-BNLM simultaneously an additional 1\% of WERR can be obtained.

\subsubsection{Extended word-based augmentation}

\begin{figure}[tbp]
\centering
\includegraphics[width=1.0\columnwidth]{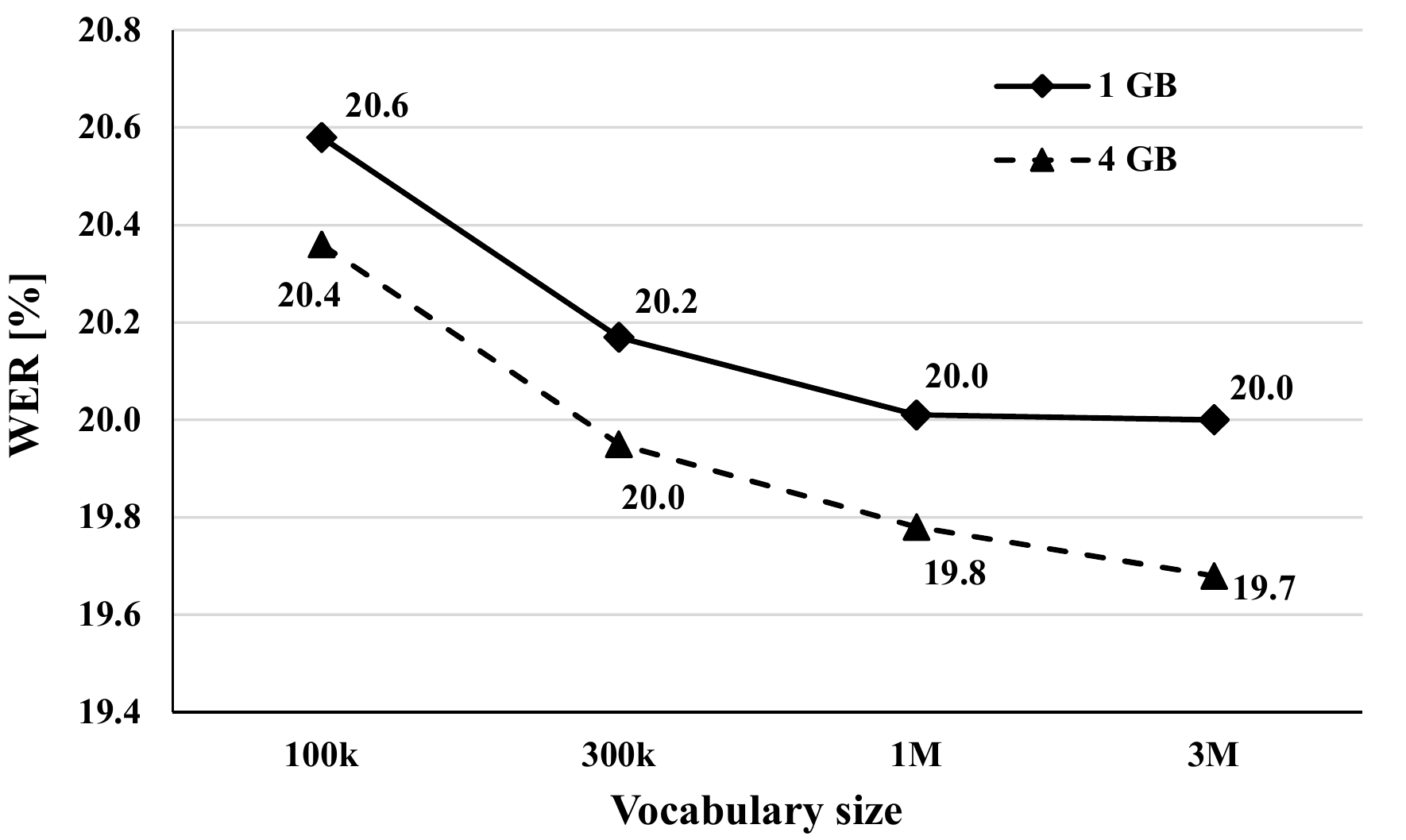}
\caption{WER of word-based BNLM+TR-BNLM with extended vocabulary and memory footprint}
\label{fig:wer_word_ext}
\end{figure}

In the previous section, we limited the vocabulary size of language models to 100k and pruned them to a maximum memory footprint of 1~GB for comparability reasons.
In the following, we examine the performance of word-based augmented models without these limitations (See Fig.~\ref{fig:wer_word_ext}).

As it can be seen in a morphologically rich language like Hungarian, the 100k vocabulary size is a strict limitation.
By increasing the vocabulary size to 300k, we can reduce the WER by a relative 2\% (from 20.6\% to 20.2\%) and by raising it to 1M by a relative 3\% (from 20.6\% to 20.0\%).
If we reduce LM pruning and let the memory footprint to increase from 1~GB to 4~GB, the WERR can go up to 4.5\% (WER from 20.6\% to 19.7\%), but for such a great improvement we need an extremely large vocabulary with 3 million words.
We can see that in a morphologically rich language, exploiting full advantages of neural text generation based data augmentation sacrifices footprint, as large vocabulary and high memory consumption are produced, which severely limits the practical applicability of the approach.

\subsection{Results with subword-based augmentation}

In order to lower the resource requirements of the augmented language model and utilize the generated text more efficiently, we apply subword LMs (see Sec.~\ref{sec:subword_aug}).
While in the word-based case the OOV ratio is around 1\% even with an extremely large 3-million-word vocabulary, the subword-based augmented language models (Subword BNLM + TR-BNLM) can fully cover the test set (0\% OOV ratio) with only 40k vocabulary items. 

As shown in Table~\ref{table:wer_word_morph}, the Morfessor tokenizer slightly outperforms BPE algorithm, however the difference is not statistically significant.
Subword modeling can reduce the WER of the 100k word-based model by up to 5\% (from 20.6\% to 19.6\%).
The subword BNLM + TR-BNLM, moreover, outperforms the 3-million-word vocabulary word-based model by reducing WER by 2\% relative (from 20.0\% to 19.6\%).
Both former improvements were found statistically significant (p \textless~0.05).
The WER of the subword-based model with 40k vocabulary and 1~GB memory consumption is roughly the same as the WER of the word-based model with 3M vocabulary items and 4~GB memory usage (19.6\% WER vs. 19.7\%).
Thus, we can state that neural text generation based data augmentation with subword tokenization can be significantly more efficient than word-based augmentation for a morphologically rich ASR task.

Just like in the case of word-based models (see Sec.~\ref{label:results_word_100k}), using the text generated by the recurrent model, we were able to achieve an additional average WER reduction of 1\%.

\begin{table}[tbp]
\centering
\caption{WER of word and subword-based augmentation with normal (1 GB) and extended memory footprint (4 GB)}
\resizebox{\columnwidth}{!}{%
\begin{tabular}{lllcc}
\hline
\multicolumn{2}{l}{\multirow{2}{*}{\textbf{Model}}}                                                & \multirow{2}{*}{\textbf{\begin{tabular}[c]{@{}l@{}}Vocab\\size\end{tabular}}} & \multicolumn{2}{c}{\textbf{WER {[}\%{]}}} \\
\multicolumn{2}{l}{}                                                                               &                                                                                 & \textbf{1 GB}       & \textbf{4 GB}       \\ \hline
\multicolumn{2}{l}{\multirow{4}{*}{\begin{tabular}[c]{@{}l@{}}Word\\ BNLM + TR-BNLM\end{tabular}}} & 100k                                                                            & 20.6                & 20.4                \\
\multicolumn{2}{l}{}                                                                               & 300k                                                                            & 20.2                & 20.0                \\
\multicolumn{2}{l}{}                                                                               & 1M                                                                              & 20.0                & 19.8                \\
\multicolumn{2}{l}{}                                                                               & 3M                                                                              & 20.0                & 19.7                \\ \hline
\multicolumn{2}{l}{\begin{tabular}[c]{@{}l@{}}Subword -- BPE\\ BNLM + TR-BNLM\end{tabular}}               & 40k                                                                             & 19.7                & 19.4                \\ \hline
\multicolumn{2}{l}{\begin{tabular}[c]{@{}l@{}}Subword -- Morfessor\\ BNLM + TR-BNLM\end{tabular}}               & 40k                                                                             & 19.6                & 19.3                \\ \hline
\multicolumn{2}{l}{\begin{tabular}[c]{@{}l@{}}Subword -- Morfessor\\ BNLM + TR-BNLM + RNN-BNLM\end{tabular}}    & 40k                                                                             & 19.4                & 19.1                \\ \hline
\end{tabular}%
}
\label{table:wer_word_morph}
\end{table}


\subsection{OOV recognition analysis}

The Transformer LM applied in our experiments use BPE tokenization, so it can create new word forms when generating text for data augmentation.
Thus, augmented language models become to some extent capable of recognizing Out Of Vocabulary (OOV) words, as well.
In this section, we compare this OOV recognition capability of the augmentation approaches.
We consider OOV words to be those words that did not occur in the original in-domain training text (see Sec~\ref{database}).
We evaluated the ASR outputs of word and subword-based augmentation approaches using information retrieval metrics (Precision, Recall, F\textsubscript{1})~\cite{fawcett2006introduction}.

\begin{figure}[tbp]
\centering
\includegraphics[width=1.0\columnwidth]{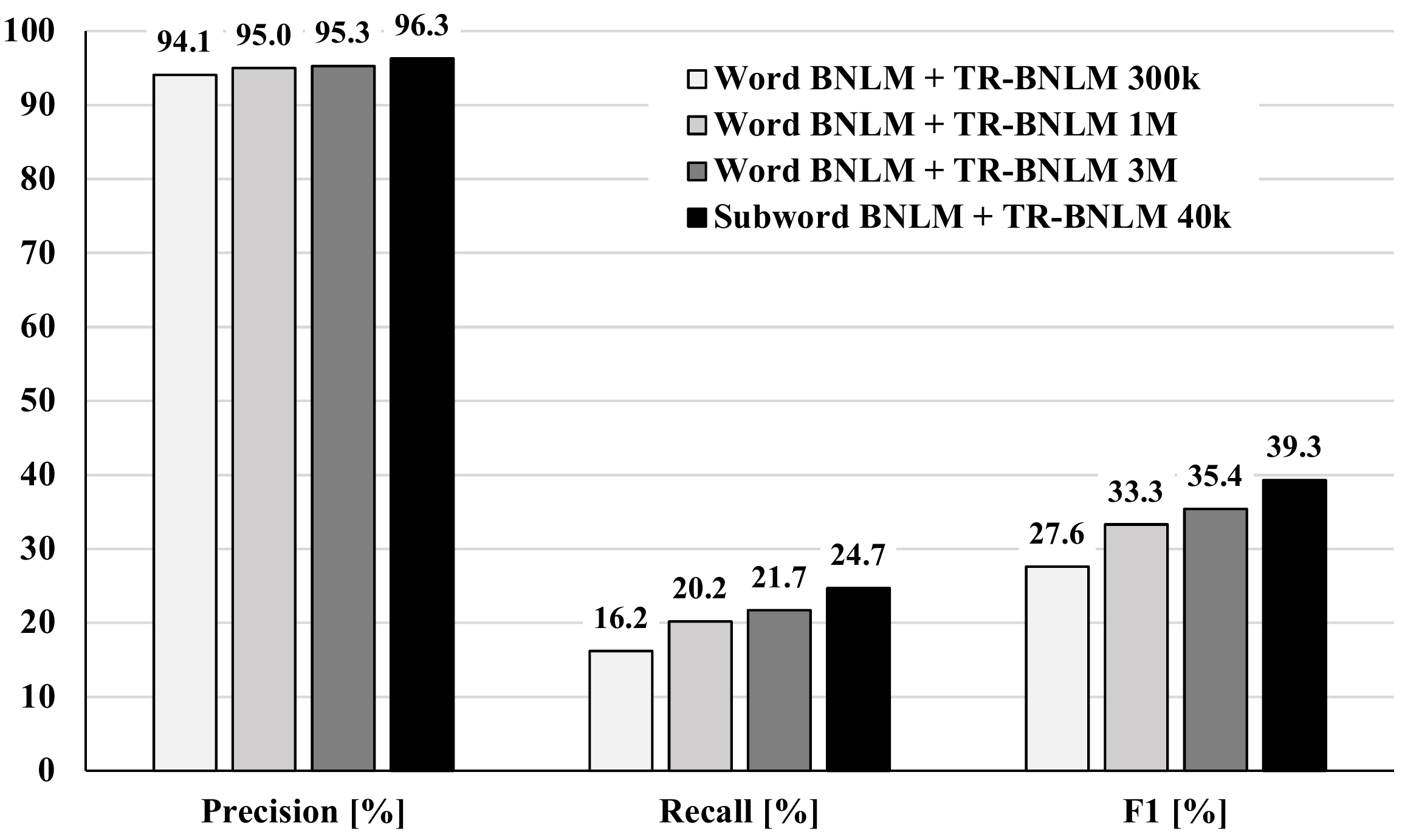}
\caption{Precision, recall and F\textsubscript{1} of OOV word recognition with various augmented language models}
\label{fig:oov}
\end{figure}

The results are summarized in Fig.~\ref{fig:oov}.
The baseline BNLM and the word-based BNLM + TR-BNLM 100k models are not shown in the figure, since they are (obviously) not capable of recognizing OOV words.
As it can be seen in Fig.~\ref{fig:oov}, all models recognize OOVs with high precision, so it is not typical that OOV words get inserted or replace other words in the ASR transcript.
What shows a significant difference between the systems examined is the value of the recall.
As the vocabulary size of word-based models increases, so does the recall of OOV words.
The 3-million-word vocabulary word-based augmented LM is capable of recognizing almost 22\% of OOVs.
However, the subword-based system can capture every 4th OOV word (\tildeapprox 25\% recall) with only 40k subwords in its vocabulary.

\section{Conclusions} \label{conclusions}

We introduced an approach called \textit{subword-based neural text augmentation} that is the extension of the Transformer based LM augmentation method presented in~\cite{Wang2019a} for morphologically rich languages.
With this new approach we managed to improve the WER of our online ASR system on Hungarian call center conversions by more than 10\% relative (from 21.9\% to 19.6\%).
Our solution also outperforms the original, word-based data augmentation technique in terms of WER and OOV recognition capability while keeping the vocabulary size and memory requirements of the system quite low.
Besides, to the best of our knowledge this is the first paper applying GPT-2 Transformer to generate augmentation data for an ASR language model.


\section{Acknowledgements}

The research was supported by the CAMEP (2018-2.1.3-EUREKA-2018-00014) and NKFIH FK-124413 projects.

\bibliographystyle{IEEEbib}
\bibliography{SLT_2021}

\end{document}